\documentclass[12pt]{article}
\usepackage{pic03}
\usepackage{hyperref}
\usepackage{url}
\usepackage{graphicx}

\begin{document}

\title{\bf RADIATIVE EVENTS IN DIS OF UNPOLARIZED ELECTRON BY
TENSOR POLARIZED DEUTERON. RADIATIVE CORRECTIONS.}
\author{
G.I Gakh      \\ {\em NSC "Kharkov Institute of Physics and
Technology",} \\ { \em Akademicheskaya 1,  Kharkov, 61108,
Ukraine} \\ O. Shekhovtsova      \\ {\em NSC "Kharkov Institute of
Physics and Technology",} \\ { \em  Akademicheskaya 1,  Kharkov,
61108, Ukraine}} \maketitle \baselineskip=17pt

One of the main objective of the HERA experiments is determination
of structure functions (SF) of the nucleons over broad range of
the kinematic variables. For some purposes it is necessary to
measure the deep--inelastic scattering (DIS) cross section at
different energies but running  the collider at reduced beam
energies increases some systematic errors. To solve this problem
it was suggested  to use the radiative events \cite{KPS}.

To extract information on the neutron spin--dependent SF, $g_1(x)$
polarized deuterons are used. However, the polarized deuteron is
interesting in its own, because it has spin one: there are
additional spin--dependent SF (caused by deuteron tensor
polarization) \cite{HST}.


The spin--dependent part of the DIS of unpolarized electron beam
from the tensor polarized deuteron target was calculated, which is
accompanied by emission of the collinear hard photon
$(\theta_{\gamma} = {\bf \widehat{p_1k}} \leq\theta_0, \
\theta_0\ll 1),$
\begin{equation}\label{1,radiative process}
e^-(p_1) +d^T(p) \rightarrow e^-(p_2) + \gamma(k) + X,
\end{equation}
(the spin--independent part of the DIS cross section with tagged
photon has been estimated in Ref. \cite{GKM}).

In the Born approximation the DIS cross section of the process (1)
can be written as
\begin{equation}\label{10,born cross section}
\frac{z}{y}\frac{d\sigma^{B}}{dxdydz}= \frac{2\pi\alpha^2(\hat
Q^2)}{\hat y^2\hat Q^4}\frac{\alpha}{2\pi}P(z,L_0)
\bigl[S_{ll}Q_{ll} +S_{tt}(Q_{tt}-Q_{nn})+S_{lt}Q_{lt}\bigr],
\end{equation}
where $x=\displaystyle\frac{\mathstrut Q^2}{2p(p_1-p_2)}$,
$y=\displaystyle\frac{\mathstrut 2p(p_1-p_2)}{V}$, $V=2pp_1$,
$\hat Q^2=zQ^2$. The variable $z=2p(p_1-k)/V$ is the energy
fraction of the electron after the initial state radiation of a
collinear photon. The function $P(z, L_o)$ has the following form
\begin{equation}
P(z,L_0)=\frac{1+z^2}{1-z}L_0-\frac{2z}{1-z} \ , \ \ \
L_0=\ln{\frac{E_1^2\theta_0^2}{m^2}},
\end{equation}
where $m$ is the electron mass, $E_1$ is the electron beam energy.
The functions $S_{ik}$ ($i,k=l,t,n$) depend, in the general, on
four tensor SF's $b_1, b_2, b_3, b_4$ and on the kinematical
variables also \cite{UFGh}. The quantities $Q_{ll}$, $Q_{tt}$ and
$Q_{nn}$ are the components of the deuteron quadrupole
polarization tensor in the laboratory system, where the $l$ axis
is directed along the electron beam momentum, the $t$ axis is
perpendicular to the $l$ axis and lies in the reaction plane, and
the $n$ axis is ortogonal to the reaction plane. The components
obey the relations $Q_{ij}=Q_{ji}$ and $Q_{ll}+Q_{tt}+Q_{nn}=0$.
To give an idea of the magnitude of the quantities $S_{ik}$ we
calculate them neglecting the high--twist $b_3$ and $b_4$
functions and suppose the relation $b_2=2xb_1$.

We present in Fig.1 the behavior of $S_{ll}$ and $S_{tt}$ as
function of  $y$ variable (at fixed $x$ and $z$ variables) for the
parametrization of SF $b_1$ given in Ref. \cite{Edelm}.

\begin{figure}[htbp]
  \centerline{\hbox{ \hspace{0.2cm}
    \includegraphics[width=6.5cm]{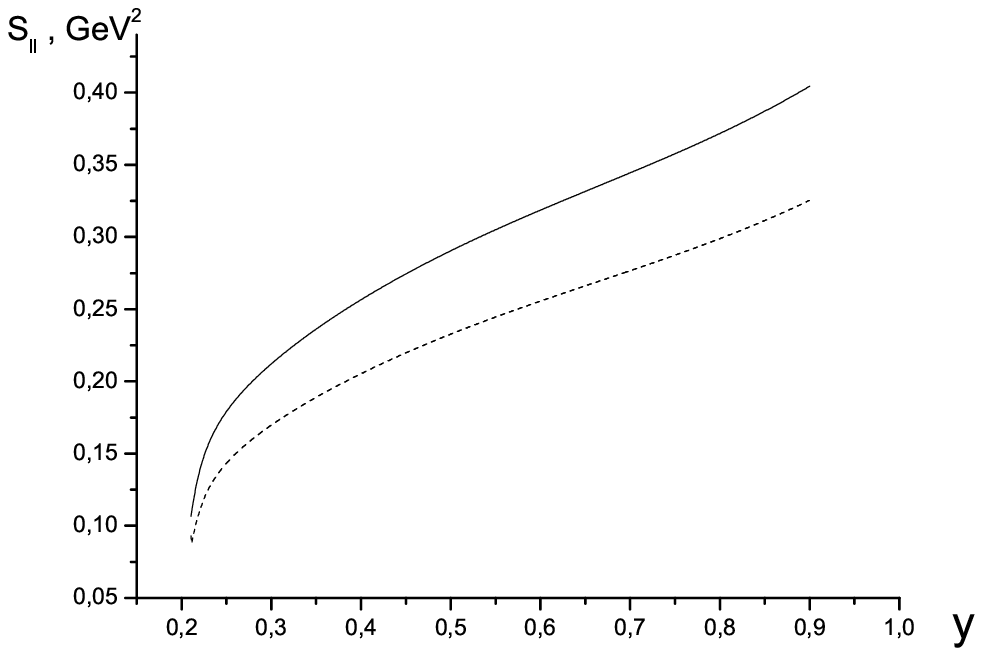}
    \hspace{0.3cm}
   \includegraphics[width=6.5cm]{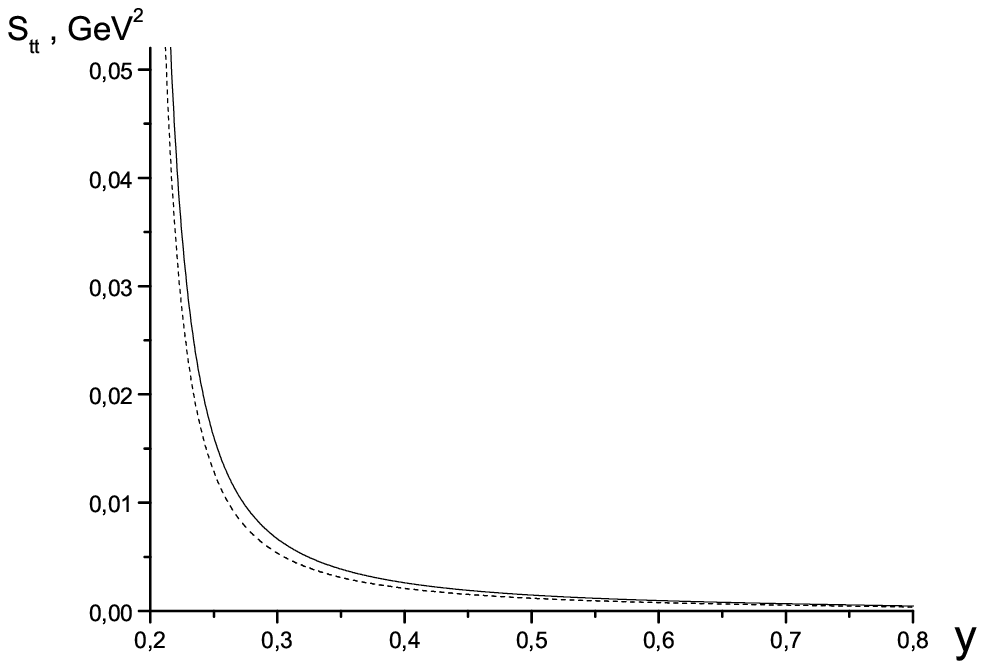}
    }
  }
 \caption{\it
      $y$--dependence of the quantities
 $\hat{S_{ll}}$, $\hat{S_{tt}}$
 for $x=0.1$ and $z=0.82$ (this choice corresponds to
 the photon energy of 5 GeV, the beam energy is 27.6 GeV). The solid and dash lines  correspond to the Paris
 and Bonn potentials for deuteron wave functions, respectively.
    \label{exfig} }
\end{figure}

We have also calculated QED radiative corrections (RC) to the
process (1) \cite{UFGh}. The differential cross section of the
reaction (1), taking into account  QED RC, can be written as
\begin{equation}
\frac{d\sigma}{dxdydz}=\frac{d\sigma^{B}}{dxdydz}(1+\delta),
\end{equation}
where $\delta$--term  is determined by RC. We calculated RC for
the cases of the exclusive and calorimeter event selections
\cite{UFGh}. $y$--dependence of the Born cross section and
$\delta$ (under the calorimeter event selection, for
$\theta_0=0.5mrad$, $\theta'=50mrad$) is shown in Fig.2 for the
same kinematical conditions as in Fig.1. For the polarization
state of the target we used the target parameters of the Ref.
\cite{HERMES}. The tensor polarization of the deuteron target is
determined by the quantity $T=1-3n_0$, where $n_0$ is the atomic
population with zero spin projection onto the quantization axis.
We consider the case when the quantization axis is directed along
the momentum of the initial electron.

\begin{figure}[htbp]
  \centerline{\hbox{ \hspace{0.2cm}
  \includegraphics[width=6.5cm]{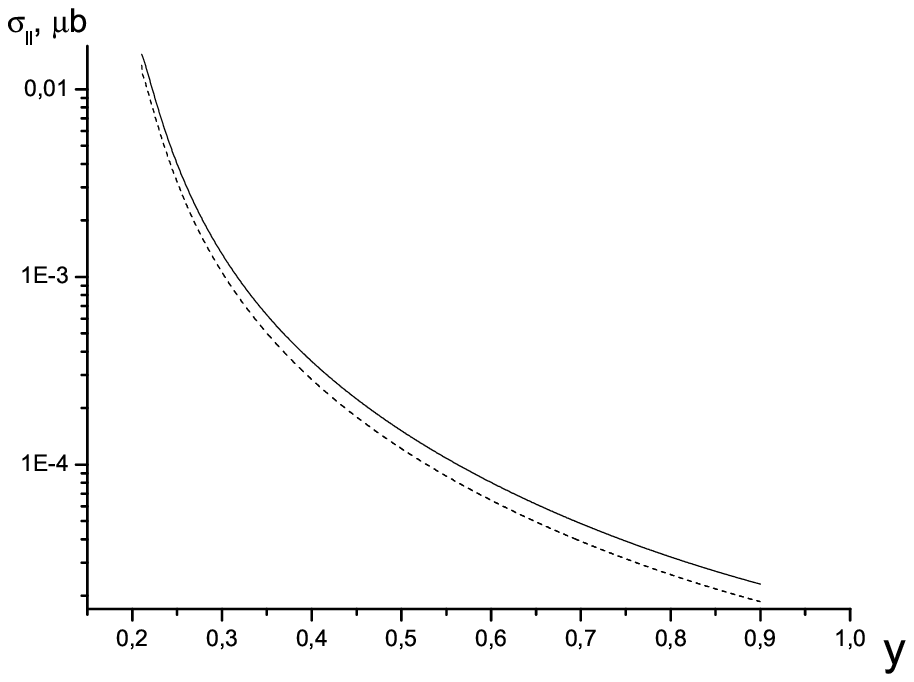}
    \hspace{0.3cm}
    \includegraphics[width=6.5cm]{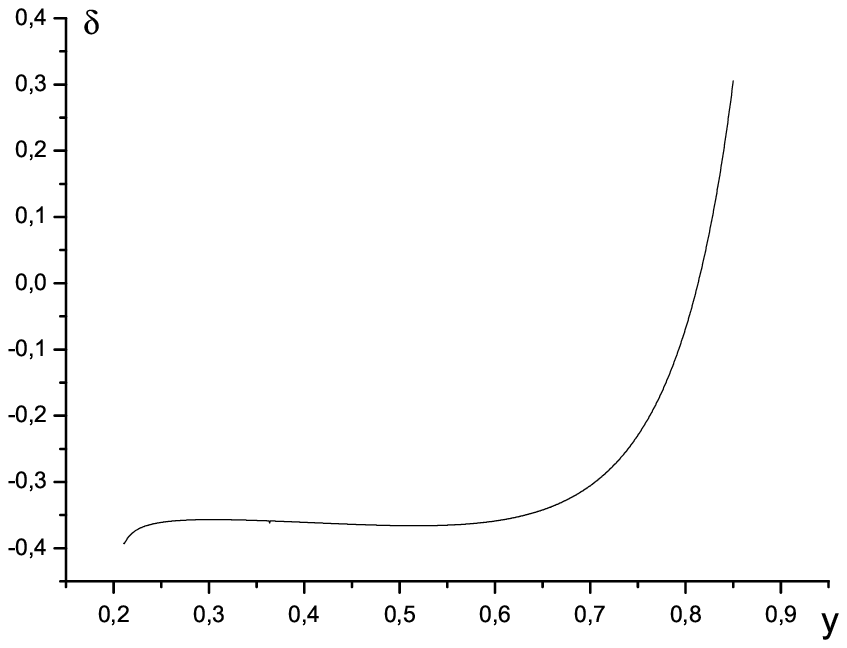}
    }
  }
 \caption{\it
     $y$--dependence of the cross section (2) (the left figure)
  and the quantity $\delta$  (the right figure) in the case of
  the longitudinal  polarization for the target polarization $T=0.83$
   \cite{HERMES}. The rest notations are the same as Fig.1}
    \label{figsect}
\end{figure}

\section{Acknowledgements}
We acknowledge numerous usful discussions with Prof. N.P.Merenkov.

\end{document}